\documentclass[pdflatex,sn-mathphys-num]{sn-jnl}

\usepackage{graphicx}       
\usepackage{amsmath}        
\usepackage{amssymb}        
\usepackage{physics}        
\usepackage{xcolor}         
\usepackage{url}            
\usepackage{hyperref}
\usepackage{bbold}

\hypersetup{
  colorlinks=true,
  linkcolor=blue,
  citecolor=blue,
  urlcolor=blue
}

\title{Strain-Induced Detuning of a Dressed Nitrogen-Vacancy Qubit: Effective Two-Level Theory and Its Validity}

\author[1]{\fnm{Jihyeon} \sur{Jun}}
\author[2]{\fnm{Donghun} \sur{Lee}}
\author[3]{\fnm{Seok-Kyun} \sur{Son}}
\author*[1,4]{\fnm{Nojoon} \sur{Myoung}}\email{nmyoung@chosun.ac.kr}

\affil[1]{\orgdiv{Department of Physics Education},
           \orgname{Chosun University},
           \orgaddress{
             \city{Gwangju},
             \postcode{61452},
             \country{Republic of Korea}}}

\affil[2]{\orgdiv{Department of Physics},
          \orgname{Korea University},
          \orgaddress{
            \city{Seoul},
            \postcode{02841},
            \country{Republic of Korea}}}

\affil[3]{\orgdiv{Department of Physics},
          \orgname{Kyung Hee University},
          \orgaddress{
            \city{Seoul},
            \postcode{02447},
            \country{Republic of Korea}}}

\affil[4]{\orgdiv{Institute of Well-Aging Medicare, CSU G-Lamp Project Group},
           \orgname{Chosun University},
           \orgaddress{
             \city{Gwangju},
             \postcode{61452},
             \country{Republic of Korea}}}

\abstract{
The nitrogen-vacancy (NV) center in diamond can be operated as a microwave-dressed qubit. In the ideal two-level limit, its transition frequency is first-order insensitive to static magnetic fields, providing robustness against magnetic detuning noise. In practical diamond devices, however, residual transverse crystal strain mixes the $\ket{m_{s}=\pm1}$ spin sublevels and modifies the dressed qubit. In this study, we derive an analytical effective two-level model of a strained dressed NV qubit by perturbatively eliminating the far-detuned spectator state from the full three-level dressed Hamiltonian. We obtain closed-form expressions for the dressed-state splitting, the spin-locking mixing angle, and the longitudinal magnetic-field coupling. We show that transverse strain shifts the dressed-state resonance and tilts the spin-locking axis. These two effects restore a finite DC-field response and thereby quantify the loss of magnetic robustness. We demonstrate these features in simulated pulsed electron spin resonance spectra that incorporate rate-equation-based optical readout. We further derive exact validity criteria from the eigenvalues and spectator weights of the full three-level Hamiltonian. For practical use, we reduce these criteria to two controlled guidelines: the spectator-like branch must remain above the nominal upper dressed state, and its branch-specific admixture must remain small. A validity diagram over the axial-field--transverse-strain plane summarizes these approximate conditions and provides practical guidelines for designing dressed-NV sensing experiments.
}


\begin{document}

\maketitle

\section{Introduction}\label{sec:intro}

The negatively charged nitrogen-vacancy (NV) center in diamond is a leading solid-state platform for quantum sensing and quantum information. Its spin-1 ground state can be optically initialized and read out, and can be coherently manipulated by microwave (MW) fields at room temperature\cite{steiner2010universal,dobrovitski2013quantum,chen2015subdiffraction,du2024single,fischer2025spin}. In the detection of static and low-frequency signals, however, slow fluctuations of the spin-transition frequency can limit the available interrogation time\cite{dolde2014high,pfender2017nonvolatile,barry2020sensitivity,delord2020spin,poulsen2022optimal,hermann2024extending,orphal2025coherent}. Continuous dynamical decoupling provides a way to suppress this detuning noise. A resonant MW dressing field with Rabi frequency $\Omega$ transforms a selected bare spin transition into a \textit{dressed} qubit. In the ideal two-level limit, the dressed splitting is determined by $\Omega$ and is first-order insensitive to slow fluctuations of the bare transition\cite{cai2012robust,xu2012coherence,macquarrie2015continuous}. The resulting increase in the usable interrogation time can improve the sensitivity to DC and slowly varying perturbations, making the dressed NV qubit an attractive sensing platform\cite{gao2026,salhov2024,aiello2013,wang2022,wang2021}.

Real NV centers, however, are generally subjected to transverse perturbations. In dense ensembles, nanodiamonds, and strain-engineered or nanofabricated devices, appreciable transverse crystal strain can introduce an off-diagonal coupling between the $m_{s}=\pm1$ sublevels\cite{liu2026,delord2025,rathi2026,du2024single,barry2020sensitivity,udvarhelyi2018}. Under MW driving of the $\ket{0}\leftrightarrow\ket{-1}$ transition, this strain-induced mixing introduces a coupling to a spectator state that is not removed by the dressing field. The resulting spectator-state admixture modifies the dressed-qubit splitting and rotates the spin-locking axis. Nevertheless, most analyses of dressed NV qubits employ an idealized two-level description\cite{gao2026,salhov2024,stark2017,louzon2025,du2024single}. To our knowledge, a quantitative analytical treatment of the strain-modified dressed eigenstates and splitting, together with clear validity criteria for the reduced two-level model, has not yet been established.

In this study, we address these two issues. Starting from the exact three-level Hamiltonian of the dressed NV ground state, we eliminate the off-resonant spectator level and derive a closed-form effective two-level Hamiltonian. This model provides explicit expressions for the dressed splitting, the spin-locking mixing angle, and the longitudinal-probe matrix element. We show that transverse strain shifts the dressed-state resonance and tilts the spin-locking axis. These effects restore a finite DC-field response and provide a quantitative measure of the lost magnetic robustness. We demonstrate the resulting spectral signatures by simulating pulsed projective electron spin resonance (ESR) spectra with rate-equation-based photoluminescence (PL) readout. Finally, we determine the validity range of the reduced description from the exact eigenvalues and spectator weights of the full Hamiltonian. We then obtain a transparent operating guideline by approximating the exact boundaries in terms of the diabatic spectator--upper-dressed detuning and the corresponding mixing ratio. The resulting validity diagram distinguishes the region in which the analytical reduction is accurate, the region in which the full three-level model is required, and the region in which the spectator-like branch is interleaved between the two qubit-character branches.

The remainder of the paper is organized as follows. Section \ref{sec:model} develops the effective theory of the strain-dressed NV qubit. Section \ref{sec:strain} benchmarks the analytical theory against the full three-level calculation and presents the resulting pulsed-ESR signatures. Section \ref{sec:validity} constructs the practical validity diagram. The appendices provide the detailed reduction, exact branch criteria, dark-phase dynamics, and optical rate equations. Section \ref{sec:summary} summarizes the main results.

\section{Effective Theory of the Strain-Dressed NV Qubit}\label{sec:model}

\subsection{Driven Hamiltonian of the strained NV center}

The NV ground state is a spin triplet ($S=1$) with a zero-field splitting $D\simeq 2.87\,\mathrm{GHz}$ between the $m_{s}=0$ and $m_{s}=\pm1$ manifolds. We consider an axial magnetic field $B_{z}$ applied along the NV symmetry axis and a transverse strain perturbation with coupling strength $g_{\perp}$. We restrict the MW-driven spin dynamics to the three sublevels $\{\ket{+1},\ket{0},\ket{-1}\}$. In the absence of MW driving, the ground-state Hamiltonian in this basis is
\begin{align}
    \frac{H_{0}}{\hbar}=\begin{pmatrix} D+\gamma_e B_z & 0 & g_\perp\\ 0 & 0 & 0\\ g_\perp & 0 & D-\gamma_e B_z\end{pmatrix},\label{eq:h0}
\end{align}
where $\gamma_{e}=2.8\,\mathrm{MHz/G}$ is the gyromagnetic ratio\cite{liu2026,kollarics2024,apra2025,ovartchaiyapong2014dynamic,meesala2016enhanced,fuchs2008excited}. The terms $\pm\gamma_{e}B_{z}$ represent the Zeeman shifts of the $\ket{\pm1}$ sublevels, whereas $g_{\perp}$ couples them through transverse strain. The eigenfrequencies of $H_{0}$ are $\omega_{0}=0$ and $\omega_{\pm}=D\pm \sqrt{\left(\gamma_{e}B_{z}\right)^{2}+g_{\perp}^{2}}$.

\begin{figure}[thpb!]
    \centering
    \includegraphics[width=0.9\linewidth]{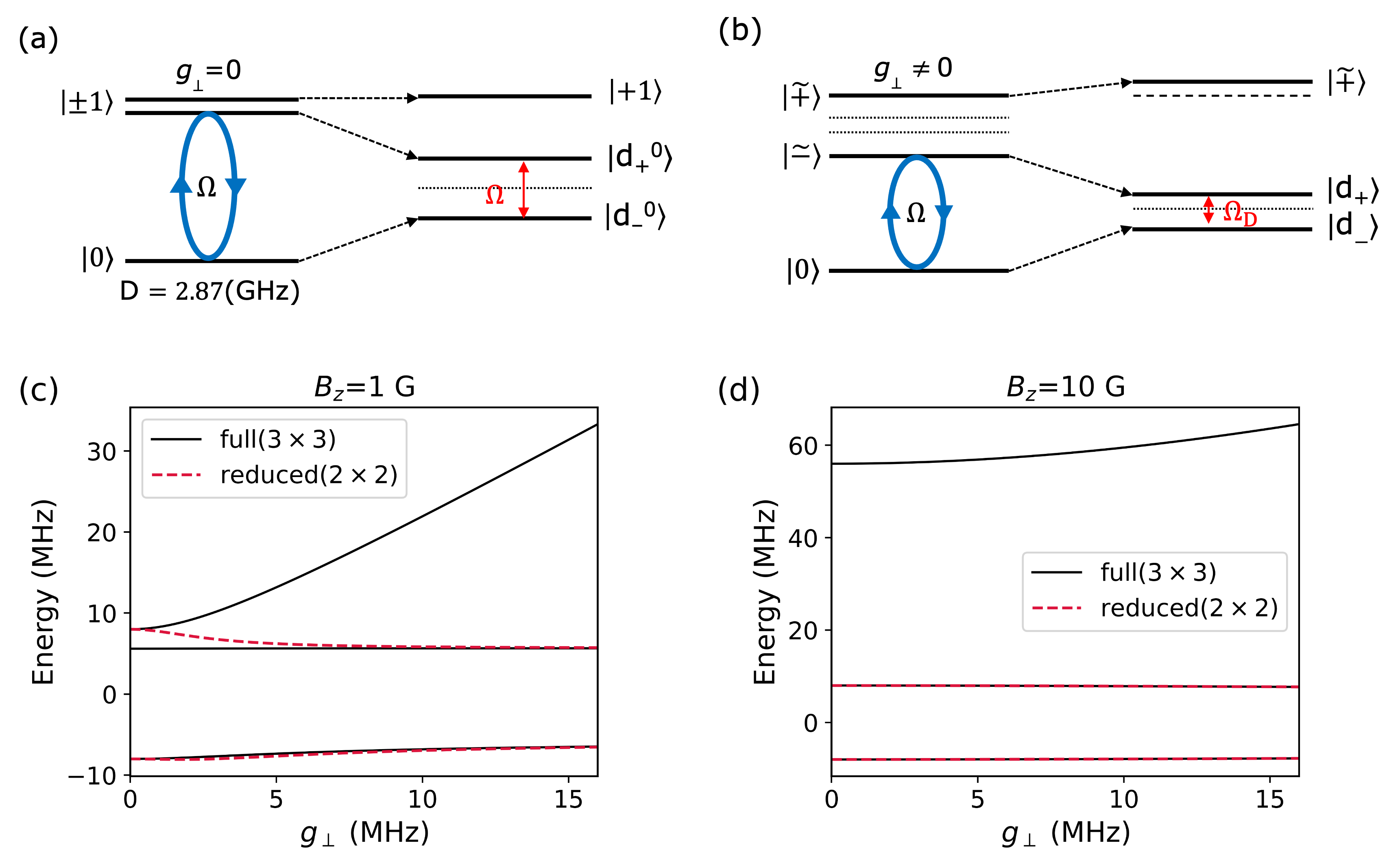}
    \caption{Dressed NV ground-state qubit under transverse strain. (a,b) Schematic dressed-level structures (a) without and (b) with strain. A resonant microwave (blue) dresses the $\ket0\leftrightarrow\ket{-1}$ transition with Rabi frequency $\Omega$. Transverse strain reduces the dressed splitting to $\Omega_D$. (c,d) Eigenfrequencies obtained from the full three-level dressed Hamiltonian [full ($3\times3$), solid] and the reduced two-level model [reduced ($2\times2$), dashed] as functions of $g_{\perp}$ at (c) $B_z=1\,\mathrm{G}$ and (d) $B_z=10\,\mathrm{G}$. The reduced levels follow the full spectrum in the strong-field regime. In the weak-field regime, the two results deviate as the spectator branch (upper solid) approaches the qubit doublet.}
    \label{fig:model}
\end{figure}

We introduce MW driving with Rabi frequency $\Omega$ and frequency $\omega_{d}=\omega_{-}$. The drive is taken to couple to the bare $\ket{0}\leftrightarrow\ket{-1}$ transition. The resulting Hamiltonian is
\begin{align}
    \frac{H_{\mathrm{full}}}{\hbar}=\frac{H_{0}}{\hbar}+\Omega\cos{\omega_{d}t}\left(\ket{0}\bra{-1}+\ket{-1}\bra{0}\right).\label{eq:ht}
\end{align}
The transverse strain in Eq. \eqref{eq:h0} mixes the $\ket{+1}$ and $\ket{-1}$ sublevels. Diagonalizing the coupled $m_{s}=\pm1$ block defines the strain eigenstates
\begin{align}
    \ket{\widetilde{-}}=\cos{\psi}\ket{-1}-\sin{\psi}\ket{+1},\qquad \ket{\widetilde{+}}=\sin{\psi}\ket{-1}+\cos{\psi}\ket{+1},
\end{align}
where the mixing angle $\psi$ satisfies
\begin{align}
    \tan{2\psi}=\frac{g_{\perp}}{\gamma_{e}B_{z}}.
\end{align}
The strain admixture causes the MW field to couple $\ket{0}$ to both strain eigenstates. The corresponding Rabi frequencies are $\Omega_{-}\equiv\Omega\cos{\psi}$ for $\ket{\widetilde{-}}$ and  $\Omega_{+}\equiv\Omega\sin{\psi}$ for the far-detuned state $\ket{\widetilde{+}}$. We transform Eq. \eqref{eq:ht} to a frame rotating at $\omega_{d}$ and apply the rotating-wave approximation. The resulting three-level dressed Hamiltonian is
\begin{align}
    \frac{H_{\mathrm{dr}}}{\hbar}=\begin{pmatrix}
        \Delta & \dfrac{\Omega_{+}}{2}&0\\
        \dfrac{\Omega_{+}}{2}&0&\dfrac{\Omega_{-}}{2}\\
        0&\dfrac{\Omega_{-}}{2}&0
    \end{pmatrix},\label{eq:hdr}
\end{align}
in the ordered basis $\{\ket{\widetilde{+}},\ket{0},\ket{\widetilde{-}}\}$, where $\Delta\equiv\omega_{+}-\omega_{-}$. In Eq. \eqref{eq:hdr}, the zero-detuning subspace spanned by $\{\ket{0},\ket{\widetilde{-}}\}$ forms the nominal dressed-qubit subspace. The state $\ket{\widetilde{+}}$ is detuned by $\Delta$ and couples to this subspace only through $\Omega_{+}$, and therefore acts as a spectator state. At zero strain ($g_{\perp}=0$), $\Omega_{+}=0$, and $\ket{\widetilde{+}}=\ket{+1}$ are completely decoupled. The ideal two-level dressed qubit is then recovered. At finite strain, $\Omega_{+}$ becomes nonzero, and the spectator state is coupled off resonance. Figures \ref{fig:model}(a) and (b) depict the dressed ground-state qubit without and with strain, respectively. The MW field dresses the $\ket{0}\leftrightarrow\ket{-1}$ transition, whereas $g_{\perp}$ reduces the dressed splitting from $\Omega$ to $\Omega_{D}$.

\subsection{Effective two-level theory}

To obtain an effective two-level Hamiltonian, we project $H_{\mathrm{dr}}$ onto the dressed-qubit subspace and eliminate the spectator state perturbatively. We denote the projector onto $\{\ket{0},\ket{\widetilde{-}}\}$ by $P$ and define $Q=\mathbb{1}-P$ as the projector onto the spectator subspace $\{\ket{\widetilde{+}}\}$. The L\"{o}wdin--Feshbach partitioning\cite{ten2013stochastic,pavlyukh2015single,shamshutdinova2008feshbach,christopher2006efficient,zhang1990photodissociation,jin2011partitioning} gives the exact energy-dependent effective Hamiltonian (see Appendix \ref{app:reduction}):
\begin{align}
    H_{\mathrm{eff}}=PH_{\mathrm{dr}}P+PH_{\mathrm{dr}}Q\left(E-QH_{\mathrm{dr}}Q\right)^{-1}QH_{\mathrm{dr}}P,\label{eq:Feshbach}
\end{align}
where $E$ denotes an eigenenergy of $H_{\mathrm{dr}}$ in the qubit sector. Because the spectator subspace is one-dimensional, $QH_{\mathrm{dr}}Q=\hbar\Delta$. In addition, $PH_{\mathrm{dr}}Q$ contains only the matrix element $\hbar\Omega_{+}/2$ between $\ket{0}$ and $\ket{\widetilde{+}}$. Equation \eqref{eq:Feshbach} therefore reduces to
\begin{align}
    \frac{H_{\mathrm{eff}}}{\hbar}=\begin{pmatrix}
        \dfrac{\Omega_{+}^{2}/4}{E/\hbar-\Delta}&\dfrac{\Omega_{-}}{2}\\
        \dfrac{\Omega_{-}}{2}&0
    \end{pmatrix},
\end{align}
in the $\{\ket{0},\ket{\widetilde{-}}\}$ basis. When the qubit frequencies are small compared with the spectator detuning, $|E|/\hbar\ll\Delta$, the energy dependence of the Feshbach resolvent can be neglected. We evaluate it at the unperturbed qubit centroid and use $E/\hbar-\Delta\approx -\Delta$. Under this condition, virtual coupling to the spectator state lowers the $\ket{0}$ level by a static AC-Stark shift with a magnitude of $\Omega_{+}^{2}/4\Delta$. The driven Hamiltonian $H_{\mathrm{dr}}$ then reduces to
\begin{align}
    \frac{H_{\mathrm{red}}}{\hbar}=\begin{pmatrix}
        \Delta_{\mathrm{eff}}&\dfrac{\Omega_{-}}{2}\\
        \dfrac{\Omega_{-}}{2}&0
    \end{pmatrix},\label{eq:hred}
\end{align}
where $\Delta_{\mathrm{eff}}\equiv-\Omega_{+}^{2}/4\Delta$. The eigenstates of this reduced Hamiltonian are the dressed-qubit states
\begin{align}
    \ket{d_{+}}=\cos{\frac{\theta}{2}}\ket{0}+\sin{\frac{\theta}{2}}\ket{\widetilde{-}},\qquad \ket{d_{-}}=-\sin{\frac{\theta}{2}}\ket{0}+\cos{\frac{\theta}{2}}\ket{\widetilde{-}}.
\end{align}
The corresponding rotating-frame eigenfrequencies are split by
\begin{align}
    \Omega_{D}^{(2)}\equiv\Omega_D=\sqrt{\Delta_{\mathrm{eff}}^{2}+\Omega_{-}^{2}},
    \label{eq:OmegaD_reduced}
\end{align}
where the spin-locking mixing angle is defined by
\begin{align}
    \tan{\theta}=\frac{\Omega_{-}}{\Delta_{\mathrm{eff}}}.
\end{align}
We choose the branch of $\theta$ continuously such that $\theta\to\pi/2$ in the unstrained limit.
The dressed-state splitting $\Omega_{D}$ determines the ESR resonant frequency. In the unstrained limit ($g_{\perp}\to0$), $\psi\to0$ and $\Delta_{\mathrm{eff}}\to0$, and hence $\Omega_{D}\to\Omega$. The system is then described by the ideal $2\times2$ dressed Hamiltonian with a resonant frequency fixed by the MW amplitude $\Omega$. At finite $g_{\perp}$, on the other hand, the dressed resonance is shifted from the bare dressed splitting.

\subsection{Analytical observables and validity scale}

The longitudinal probe couples the two dressed states through the matrix element
\begin{align}
    m_{q}=\left|\bra{d_{+}}S_{z}\ket{d_{-}}\right|=\frac{1}{2}\cos{2\psi}\sin{\theta}.
\end{align}
Within the reduced model, this matrix element decreases with strain as the axial field $B_{z}$ is lowered. It reduces the effective probe amplitude and thereby modifies the relative sideband weights in a fixed-duration pulsed spectrum.

The validity of the spectator-state elimination involves two distinct questions. First, the spectator-like branch can lie either above or between the two qubit-character branches. Second, even when the spectator remains above the nominal doublet, its admixture can make the energy-independent reduction quantitatively inaccurate. The exact distinction is determined from the eigenvalues and spectator weights of the full Hamiltonian, as derived in Appendix \ref{app:validity}. For a transparent estimate, we diagonalize the uncoupled qubit block $PH_{\mathrm{dr}}P$. Its nominal dressed states are
\begin{align}
    \ket{d_{\pm}^{(0)}}=\frac{\ket0\pm\ket{\widetilde-}}{\sqrt{2}}.
    \label{eq:nominal_dressed}
\end{align}
Their eigenfrequencies are $\pm\Omega_-/2$. The superscript $(0)$ denotes zeroth order in the spectator coupling $\Omega_+$, not zero strain; both $\ket{\widetilde-}$ and $\Omega_-$ retain their full strain dependence. In the limit $\Delta_{\mathrm{eff}}\to0$, the effective eigenstates satisfy $\ket{d_+}\to\ket{d_+^{(0)}}$ and $\ket{d_-}\to-\ket{d_-^{(0)}}$, where the minus sign is a physically irrelevant global phase. On the $\Delta>\Omega_-/2$ side, we define the branch-specific spectator-admixture parameter
\begin{align}
    \varepsilon\equiv
    \frac{|\Omega_+|/(2\sqrt{2})}{\Delta-\Omega_-/2}.
    \label{eq:epsilon}
\end{align}
The numerator $|\Omega_+|/(2\sqrt{2})$ is the coupling between the spectator and the nominal upper dressed state. The denominator $\Delta-\Omega_-/2$ is their spectral detuning. Thus, $\varepsilon$ directly measures the unwanted spectator coupling relative to the separation from the upper dressed branch. The weak-mixing condition is $\varepsilon\ll1$. This condition is not an exact boundary because it neglects the coupling-induced displacement of the three exact eigenvalues. Section \ref{sec:validity} uses $\varepsilon$ to construct a practical validity guideline, while Appendix \ref{app:validity} provides the exact classification. Figures \ref{fig:model}(c) and (d) compare the eigenvalues of the full dressed Hamiltonian $H_{\mathrm{dr}}$ and the reduced two-level Hamiltonian as functions of $g_{\perp}$ at $B_{z}=1$ and $10\,\mathrm{G}$, respectively. At $B_{z}=10\,\mathrm{G}$, the reduced doublet reproduces the full qubit levels over the entire strain range, while the spectator remains well separated. At $B_{z}=1\,\mathrm{G}$, on the other hand, the two descriptions deviate as the spectator approaches the nominal upper dressed state.

These analytical results provide the reference quantities used below. We benchmark them against the exact three-level eigenstates and calculate their pulsed-ESR signatures in Sec. \ref{sec:strain}. The exact branch extraction, dark-phase dynamics, and optical readout are given in Appendices \ref{app:validity}--\ref{app:obe}.

\section{Numerical Benchmark and Pulsed-ESR Signatures}\label{sec:strain}

We benchmark the analytical theory against the exact eigensystem of the full three-level Hamiltonian. The spectator weights are used to identify the two qubit-character branches and extract the exact splitting $\Omega_D^{(3)}$ and projected angle $\theta^{(3)}$, as detailed in Appendix \ref{app:validity}. We then calculate the pulsed-ESR spectra from the full density-matrix dynamics and convert the final populations into PL using a ten-level optical rate model (Appendices \ref{app:me} and \ref{app:obe}). Each spectrum is normalized by its own maximum PL value. The numerical results therefore benchmark the analytical frequencies and eigenstates and demonstrate their spectral signatures without addressing the absolute photon number or readout contrast.

\subsection{Strain-induced geometry and field response}

The reduced model of Sec. \ref{sec:model} shows how transverse strain enters the dressed-qubit Hamiltonian through two quantities: the projected drive amplitude $\Omega_{-}=\Omega\cos{\psi}$ and the spectator-induced Stark shift $\Delta_{\mathrm{eff}}=-\Omega_{+}^{2}/4\Delta$. We first examine how these quantities modify the spin-locking geometry and the dressed resonance. We compare the reduced quantities $\Omega_D^{(2)}$ and $\theta^{(2)}$ with the corresponding full-model quantities defined in Appendix \ref{app:validity}.

In the absence of strain ($g_{\perp}=0$), $\Delta_{\mathrm{eff}}$ vanishes, and the dressed splitting is $\Omega_{D}^{(2)}=\Omega_D^{(3)}=\Omega$. The spin-locking axis then lies in the equatorial plane of the Bloch sphere ($\theta^{(2)}=\theta^{(3)}=\pi/2$). Finite $g_{\perp}$ breaks this symmetry, tilts the spin-locking axis, and shifts the dressed resonance.

\begin{figure}
    \centering
    \includegraphics[width=0.9\linewidth]{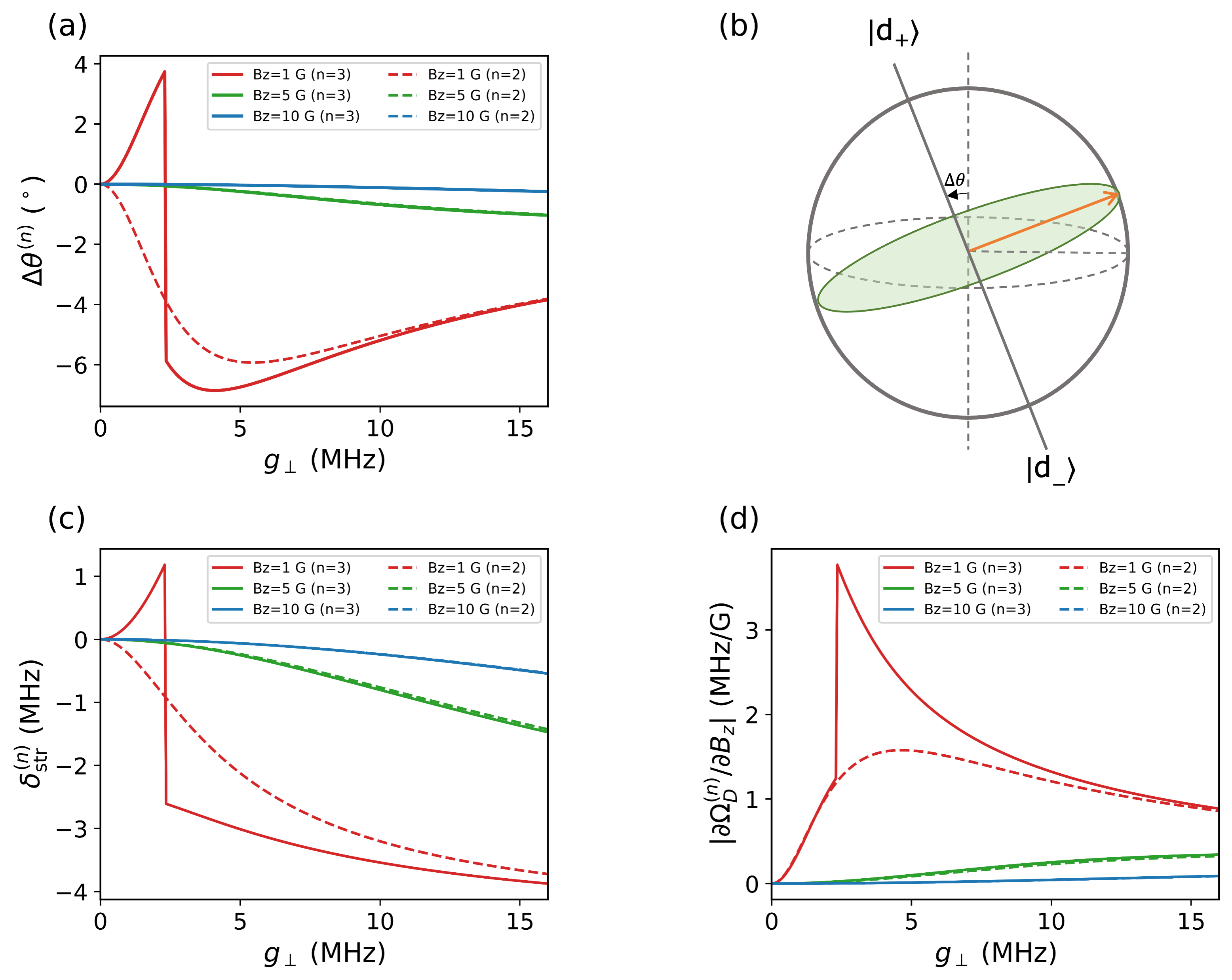}
    \caption{Strain-induced geometric distortion and residual DC-field response. (a) Projected spin-locking tilt $\Delta\theta^{(n)}=\theta^{(n)}-\pi/2$ as a function of $g_\perp$ for $B_z=1,5,10\,\mathrm{G}$ (full model, $n=3$, solid; reduced model, $n=2$, dashed). (b) Bloch-sphere representation of the strain-induced tilt. The spin-locking plane (green) and axis (orange arrow) are tilted from their unstrained orientations (dotted) by $\Delta\theta$. (c) Dressed-frequency detuning $\delta_{\mathrm{str}}^{(n)}$ and (d) residual DC-field coupling $\left|\partial\Omega_D^{(n)}/\partial B_z\right|$ as functions of $g_\perp$. These quantities vanish at $g_\perp=0$ and are largest in the weak-field regime. The discontinuous feature near $g_\perp\simeq2.5\,\mathrm{MHz}$ at $B_z=1\,\mathrm{G}$ occurs close to the diabatic spectator--upper-dressed boundary $\Delta=\Omega_-/2$. The exact eigenvalues remain continuous, but the assignment of the spectator-like branch changes across this boundary.}
    \label{fig:geometry}
\end{figure}

Within the reduced model, the spin-locking axis tilts from the equatorial plane by 
\begin{align}
    \Delta\theta^{(2)}\equiv\theta^{(2)}-\frac{\pi}{2},\qquad
    \tan{\left(\Delta\theta^{(2)}\right)}=\frac{\Delta_{\mathrm{eff}}}{\Omega_{-}},
\end{align}
as shown in Fig. \ref{fig:geometry}(a) and depicted on the Bloch sphere in Fig. \ref{fig:geometry}(b). The tilt originates from the Stark shift $\Delta_{\mathrm{eff}}\propto g_{\perp}^{2}$. It vanishes at $g_{\perp}=0$, increases with strain, and is an order of magnitude larger at $B_{z}=1\,\mathrm{G}$ than at $10\,\mathrm{G}$. Thus, the spin-locking tilt is predominantly a weak-field effect. The full-model curve is obtained from the projected angle in Eq. \eqref{eq:full_projected_angle}.

For either model, we define the signed strain-induced detuning as follows:
\begin{align}
    \delta_{\mathrm{str}}^{(n)}\equiv\Omega_{D}^{(n)}-\Omega,\qquad n=2,3.
\end{align}
Within the spectrally isolated-qubit regime, $\delta_{\mathrm{str}}^{(n)}<0$ and transverse strain redshift the dressed resonance. For the reduced model and $g_{\perp}\ll\gamma_{e}B_{z}$, the detuning reduces to $\Omega g_{\perp}^{2}/8\left(\gamma_{e}B_{z}\right)^{2}$ [Fig. \ref{fig:geometry}(c)]. In contrast to the spin-locking tilt, this redshift is dominated by the projected amplitude $\Omega_{-}=\Omega\cos{\psi}$ rather than by the Stark shift. It is second order in strain and increases toward the weak-field regime.

Together, the spin-locking tilt and resonance shift restore a finite DC-field response. In the absence of transverse strain, the dressed splitting is field independent, i.e., $\left|\partial \Omega_{D}^{(n)}/\partial B_{z}\right|=0$. Within the reduced model and at finite $g_{\perp}$, the response is approximately
\begin{align}
    \left|\frac{\partial\Omega_D^{(2)}}{\partial B_z}\right|\simeq
\frac{\Omega\,\gamma_e\,g_\perp^2}{4\left[(\gamma_e B_z)^2+g_\perp^2\right]^{3/2}},
\end{align}
as shown in Fig. \ref{fig:geometry}(d). The full-model response is obtained by differentiating $\Omega_D^{(3)}$ along the character-tracked qubit branches. This field-to-frequency transduction provides a quantitative measure of the lost robustness. The response is largest in the weak-field regime and increases as $g_{\perp}^{2}$ within the weak-strain limit.

The reduced $2\times2$ and full $3\times3$ resonance frequencies in Fig. \ref{fig:geometry} are indistinguishable at $B_z=10\,\mathrm{G}$ and remain visually close at $B_z=5\,\mathrm{G}$. At the representative point $(B_z,g_\perp)=(5\,\mathrm{G},10\,\mathrm{MHz})$, however, $\varepsilon\simeq0.064$ exceeds the conservative threshold $0.05$ used below. Thus, close agreement of a particular resonance frequency does not by itself guarantee that the spectator admixture lies below the chosen tolerance. At $B_z=1\,\mathrm{G}$, the two results deviate more strongly. The extracted full-model quantities exhibit a discontinuity near $g_{\perp}\simeq2.5\,\mathrm{MHz}$, where $\Delta\theta^{(3)}$ and $\delta_{\mathrm{str}}^{(3)}$ jump and can reverse sign. The exact eigenvalues are continuous through this anticrossing. The apparent discontinuity results from switching the eigenstate used to identify the spectator-like branch. Consequently, the assignment of $\Delta\theta^{(3)}$, $\delta_{\mathrm{str}}^{(3)}$, and $\Omega_D^{(3)}$ becomes ambiguous near the character-exchange boundary. On its interleaved side, the spectator-like state lies between the two qubit-character branches, and an energy-ordered dressed-qubit doublet is no longer spectrally isolated. Section \ref{sec:validity} gives a practical guideline for this crossover.

\subsection{Pulsed-ESR signatures}

\begin{figure*}
    \centering
    \includegraphics[width=0.9\linewidth]{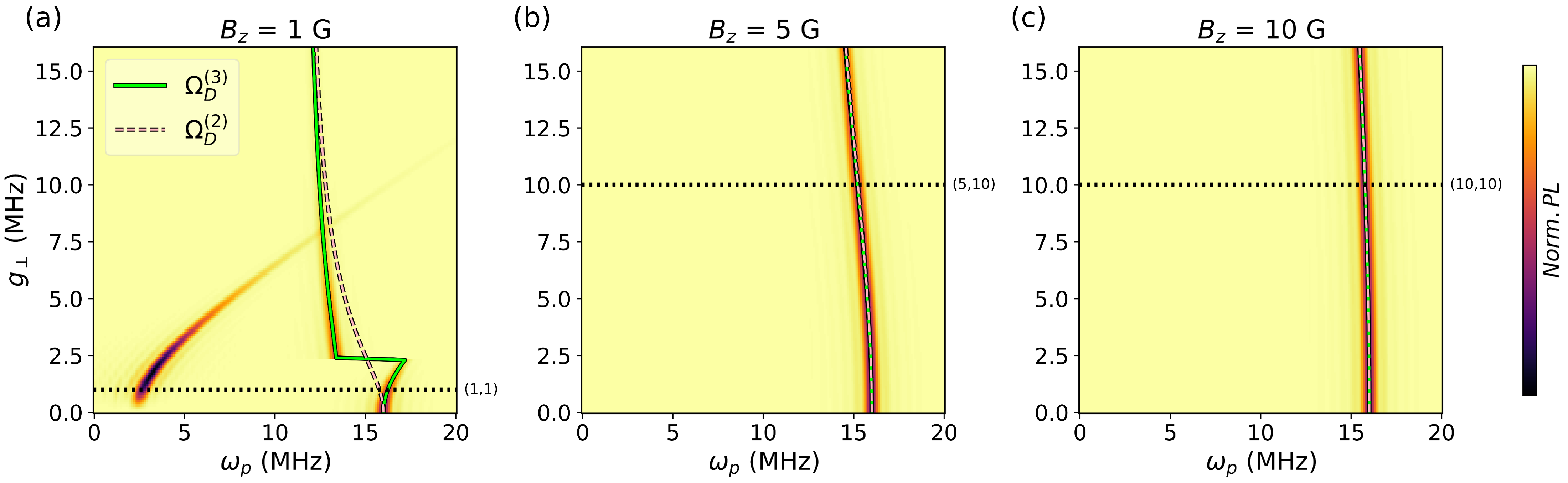}
    \caption{Pulsed-ESR spectra as functions of transverse strain. The normalized PL is plotted as a function of probe frequency $\omega_{\mathrm{p}}$ and transverse strain $g_{\perp}$ (a), (b), (c) at $B_{z}=1,5,10\,\mathrm{G}$ using the full three-level model, respectively. Solid and dashed lines indicate $\Omega_D^{(3)}$ and $\Omega_D^{(2)}$, respectively. Dotted horizontal lines denote the representative operating points of Fig. \ref{fig:phase}, labeled $(B_z,g_\perp)$. At $B_z=5$ and $10\,\mathrm{G}$, the reduced splitting follows the principal dressed-qubit resonance over the plotted strain range. At $B_z=1\,\mathrm{G}$, additional transitions involving the spectator-like branch become visible. The abrupt change in the tracked dressed-qubit trajectory near $g_\perp\simeq2.5\,\mathrm{MHz}$ reflects the spectator--qubit branch reassignment.}
    \label{fig:esr}
\end{figure*}

When the dressed-qubit transition is spectrally isolated, its pulsed-ESR dip directly measures the dressed splitting. Figure \ref{fig:esr} shows the simulated ESR signal as a function of the probe frequency $\omega_{\mathrm{p}}$ and strain $g_{\perp}$ at $B_z=1,5,$ and $10\,\mathrm{G}$. The full model also contains transitions involving the spectator-like branch, whose frequencies and intensities must be distinguished from the dressed-qubit resonance.

At $B_z=5$ and $10\,\mathrm{G}$ [Figs. \ref{fig:esr}(b) and (c)], the principal dip remains close to $\omega_{\mathrm{p}}\simeq\Omega$ and bends slightly toward lower frequencies with increasing strain. The reduced splitting $\Omega_D^{(2)}$ follows the full dressed-qubit trajectory over the plotted strain range. At $B_z=1\,\mathrm{G}$ [Fig. \ref{fig:esr}(a)], on the other hand, an additional line appears owing to a transition between the spectator-like and qubit-character branches. This transition is absent from the two-level model. The abrupt turn near $g_\perp\simeq2.5\,\mathrm{MHz}$ is the spectral counterpart of the spectator--qubit anticrossing. It marks the crossover to the region in which the spectator-like branch is interleaved between the two qubit-character branches, and the tracked dressed resonance cannot be assigned by energy ordering alone.

\section{Validity diagram} \label{sec:validity}

The reduced two-level description is not uniformly accurate over the entire parameter range. Appendix \ref{app:validity} provides an exact classification in terms of the three eigenvalues and their spectator weights. Here, we construct a compact operating guideline from controlled approximations of these exact quantities. The resulting boundaries are not singularities of the full spectrum. They identify, respectively, a change in the diabatic branch ordering and the onset of appreciable spectator mixing.

The primary guideline concerns the ordering of the spectator and nominal upper dressed states. As shown in Sec. \ref{sec:model}, the nominal upper state has eigenfrequency $\Omega_-/2$, whereas the spectator has eigenfrequency $\Delta=2\sqrt{(\gamma_eB_z)^2+g_\perp^2}$. Neglecting the coupling-induced displacement of these two branches gives us the diabatic boundary.
\begin{align}
    \Delta=\frac{\Omega_-}{2}.
    \label{eq:guideline_primary}
\end{align}
For $\Delta>\Omega_-/2$, the spectator remains above the nominal upper dressed state. For $\Delta<\Omega_-/2$, the spectator branch is diabatically interleaved between the two qubit-character branches. This classification concerns spectral ordering and does not imply that the exact three-level Hamiltonian or all possible qubit encodings cease to exist.

In the strong-field limit, $g_\perp\ll\gamma_eB_z$, one has $\Omega_-\simeq\Omega$. The primary guideline then reduces to the simple ellipse
\begin{align}
    \left(\gamma_eB_z\right)^2+g_\perp^2\simeq\left(\frac{\Omega}{4}\right)^2.
    \label{eq:primary_ellipse}
\end{align}
At $g_\perp=0$, this expression gives the threshold field
\begin{align}
    B_z^{\ast}=\frac{\Omega}{4\gamma_e}\simeq1.4~\mathrm{G}
\end{align}
for $\Omega=16~\mathrm{MHz}$. Equation \eqref{eq:primary_ellipse} is useful for visualizing the scale of the boundary, whereas Eq. \eqref{eq:guideline_primary} retains the strain dependence of the projected drive through $\Omega_- = \Omega\cos\psi$.

The secondary guideline concerns the accuracy of the two-level reduction on the $\Delta>\Omega_-/2$ side. As defined in Eq. \eqref{eq:epsilon}, the relevant ratio is
\begin{align}
    \varepsilon=
    \frac{|\Omega_+|/(2\sqrt{2})}{\Delta-\Omega_-/2}.
    \label{eq:guideline_secondary}
\end{align}
We choose the contour $\varepsilon=0.05$ as a practical boundary between regions I and II. This choice requires the spectator--upper-dressed separation to be twenty times larger than their coupling. In the weak-mixing limit, it corresponds to a spectator probability of approximately $2.5\times10^{-3}$. The exact normalized admixture and the corresponding region boundary are derived in Appendix \ref{app:validity}.

\begin{figure}
    \centering
    \includegraphics[width=0.9\linewidth]{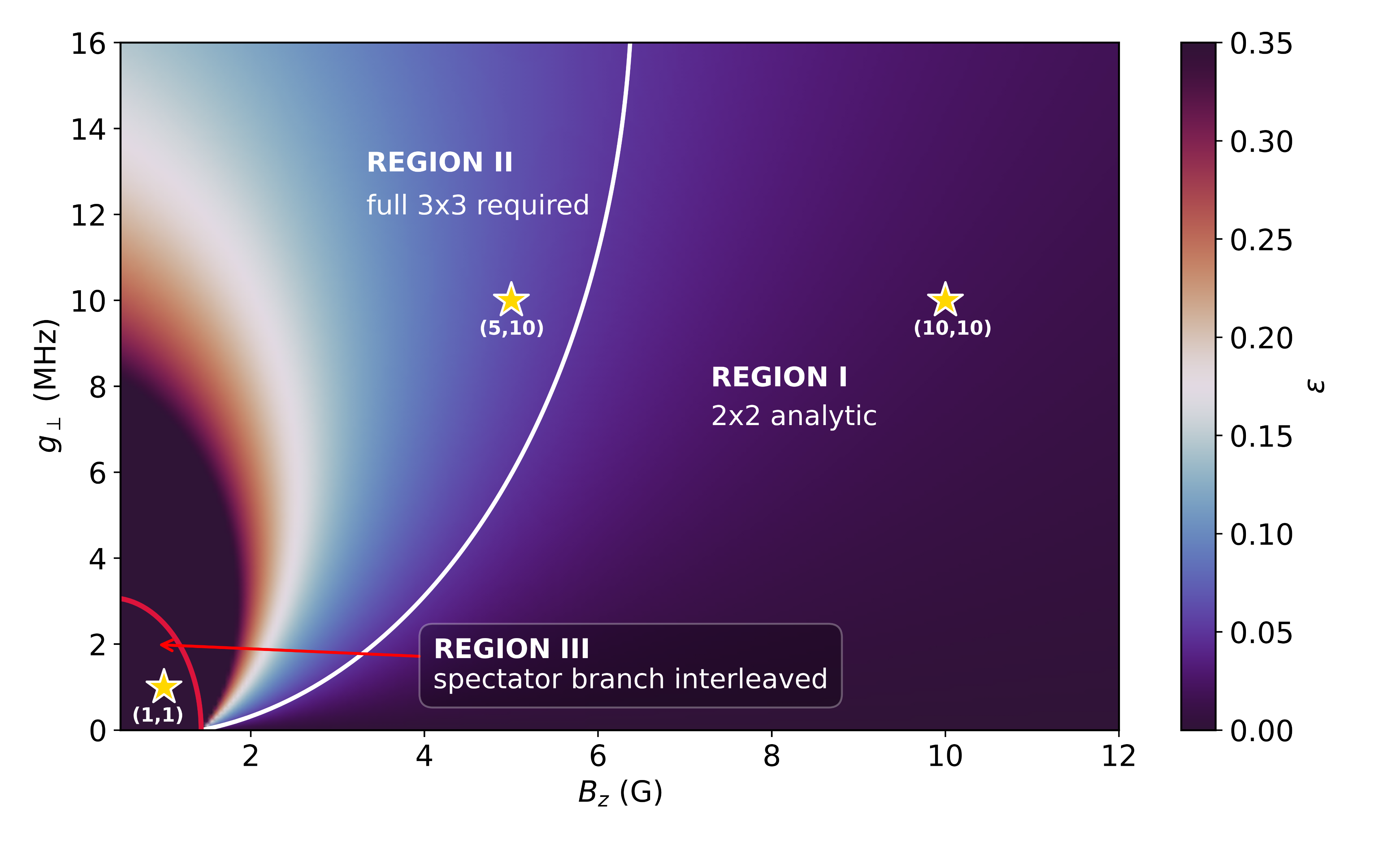}
    \caption{Approximate validity diagram of the dressed NV qubit in the $B_{z}$--$g_{\perp}$ plane for $\Omega=16\,\mathrm{MHz}$. The primary diabatic boundary $\Delta=\Omega_-/2$ (red line) separates region III, in which the spectator-like branch is interleaved between the two qubit-character branches, from regions I and II. On the $\Delta>\Omega_-/2$ side, the secondary boundary $\varepsilon=0.05$ (white line) separates region I, in which the analytical $2\times2$ reduction is accurate, from region II, in which the full $3\times3$ model is required. Here, $\varepsilon$ is the ratio of the spectator--upper-dressed coupling to their spectral detuning [Eq. \eqref{eq:epsilon}]. Stars indicate representative operating points. The plotted boundaries are practical approximations to the exact spectator-character and leakage criteria derived in Appendix \ref{app:validity}.}
    \label{fig:phase}
\end{figure}

Together, the two approximate conditions divide the $\left(B_z,g_\perp\right)$ plane into three regions, as displayed in Fig. \ref{fig:phase}. Region I satisfies $\Delta>\Omega_-/2$ and $\varepsilon<0.05$, for which the analytical reduction is expected to be accurate. Region II satisfies $\Delta>\Omega_-/2$ and $\varepsilon\geq0.05$. In this region, the spectator remains above the nominal upper dressed state, but its admixture requires the full $3\times3$ treatment. Region III satisfies $\Delta<\Omega_-/2$. Here, the spectator-like branch is interleaved with the two qubit-character branches, so an energy-ordered isolated doublet cannot be assigned without following the eigenstate character.

For $B_z=1~\mathrm{G}$ and $\Omega=16~\mathrm{MHz}$, Eq. \eqref{eq:guideline_primary} gives $g_\perp\simeq2.48~\mathrm{MHz}$, consistent with the spectral feature near $2.5~\mathrm{MHz}$ in Figs. \ref{fig:geometry} and \ref{fig:esr}. The exact equal-character criterion of Appendix \ref{app:validity} gives $g_\perp\simeq2.32~\mathrm{MHz}$, whereas the strong-field ellipse in Eq. \eqref{eq:primary_ellipse} gives $2.86~\mathrm{MHz}$. These differences quantify the successive approximations used to construct the practical diagram.

Both approximate boundaries shift with the dressing amplitude. Reducing $\Omega$ lowers the nominal upper dressed energy and decreases its coupling to the spectator. It therefore contracts region III and enlarges region I, extending the usable range toward weaker fields. This extension comes at the expense of a smaller dressed splitting and weaker noise suppression. The validity diagram thus provides a practical guideline for selecting the operating point $\left(B_z,g_\perp\right)$ and the dressing amplitude $\Omega$ for strained dressed-NV sensing.

\section{Summary and Conclusions} \label{sec:summary}

In this study, we have developed an analytical effective two-level theory of a dressed nitrogen-vacancy qubit subjected to transverse strain. Starting from the full three-level dressed Hamiltonian, we perturbatively eliminate the strain-mixed spectator state using L\"{o}wdin--Feshbach partitioning. The resulting model provides closed-form expressions for the dressed-state splitting $\Omega_{D}$, the spin-locking mixing angle $\theta$, and the longitudinal-probe matrix element $m_{q}$.

Within the parameter range in which the two qubit-character branches remain spectrally identifiable, transverse strain modifies the dressed qubit in two ways. First, the reduced drive amplitude $\Omega_-=\Omega\cos\psi$ produces a redshift of the dressed resonance. Second, the spectator-induced AC-Stark shift $\Delta_{\mathrm{eff}}$ tilts the spin-locking axis out of the equatorial plane. Together, these effects restore a finite DC-field response, $|\partial\Omega_D/\partial B_z|$, that vanishes in the unstrained limit. The reduced formulas provide a quantitative description in region I, while the full three-level calculation is required when the spectator admixture exceeds the chosen tolerance in region II. In region III, the spectator-like branch is interleaved between the two qubit-character branches. A unique energy-ordered dressed doublet, and hence a signed redshift assigned without character tracking, is then no longer well defined. We demonstrate these regimes in simulated pulsed projective ESR spectra with normalized optical rate-equation readout.

The reduced description is not valid over the entire parameter range. We derive exact boundaries from the eigenvalues and spectator weights of the full three-level Hamiltonian. We then obtain a practical validity diagram by approximating the exact spectator-character and leakage criteria with the diabatic condition $\Delta=\Omega_-/2$ and the mixing ratio $\varepsilon$, respectively. The diagram identifies the region in which the analytical model is accurate (region I), the region in which the full three-level treatment is required (region II), and the region in which the spectator-like branch is interleaved between the two qubit-character branches (region III). Reducing the dressing amplitude $\Omega$ contracts region III and extends the range in which the analytical reduction can be used.

Our results provide a practical guide for designing and interpreting dressed-NV sensing experiments in strained environments, including dense ensembles, nanodiamonds, and nanofabricated devices. Because the transverse-strain coupling $g_{\perp}$ also represents transverse electric fields\cite{lee2016strain,dolde2014nanoscale,ovartchaiyapong2014dynamic,acosta2012dynamic,poem2015broadband,udvarhelyi2018}, the same framework can be applied to strain and electric field sensing with dressed spins. The approach may also be extended to other continuously driven spin-1 defect qubits\cite{kuno2026concatenated}. 

\backmatter
\section*{Acknowledgements}

This work was
supported by the National Research Foundation of Korea (NRF) Grant funded by the Korean government (MSIT and MOE) (RS-2025-00557045, RS-2023-00285353), by the Institute for Information \& communications Technology Planning \& evaluation (IITP) by MSIT (RS-2025-25464832, RS-2025-25464788), by the Korea Research Institute for Defense Technology Planning and Advancement (KRIT)—grant funded by the Defense Acquisition Program Administration (DAPA) (KRIT-CT-23-031), and by a grant from Kyung Hee University in 2023. (KHU-202311176).

\section*{Data Availability}

The data that support the findings of this study are available from
the corresponding author upon reasonable request.

\section*{Declarations}

\textbf{Conflict of Interest}
The authors declare that they have no conflict of interest.

\textbf{Author Contributions}
J.J. performed the numerical calculations and prepared all figures. N.M. conceived the idea and established the main framework of this work. J.J. and N.M. analyzed the calculated results. D.L. and S.-K.S. reviewed the study's findings and implications. All authors contributed to writing the manuscript.

\bibliography{StrainDressedNV}

@article{fischer2025spin,
  title={Spin-photon correlations from a Purcell-enhanced diamond nitrogen-vacancy center coupled to an open microcavity},
  author={Fischer, Julius and Herrmann, Yanik and Wolfs, Cornelis FJ and Scheijen, Stijn and Ruf, Maximilian and Hanson, Ronald},
  journal={Nat. Commun.},
  year={2025},
  publisher={Nature Publishing Group UK London}
}

@article{dobrovitski2013quantum,
  title={Quantum control over single spins in diamond},
  author={Dobrovitski, VV and Fuchs, GD and Falk, AL and Santori, C and Awschalom, David D},
  journal={Annu. Rev. Condens. Matter Phys.},
  volume={4},
  number={1},
  pages={23--50},
  year={2013},
  publisher={Annual Reviews}
}

@article{du2024single,
  title={Single-molecule scale magnetic resonance spectroscopy using quantum diamond sensors},
  author={Du, Jiangfeng and Shi, Fazhan and Kong, Xi and Jelezko, Fedor and Wrachtrup, J{\"o}rg},
  journal={Rev. Mod. Phys.},
  volume={96},
  number={2},
  pages={025001},
  year={2024},
  publisher={APS}
}

@article{chen2015subdiffraction,
  title={Subdiffraction optical manipulation of the charge state of nitrogen vacancy center in diamond},
  author={Chen, Xiangdong and Zou, Changling and Gong, Zhaojun and Dong, Chunhua and Guo, Guangcan and Sun, Fangwen},
  journal={Light Sci. Appl.},
  volume={4},
  number={1},
  pages={e230--e230},
  year={2015},
  publisher={Nature Publishing Group}
}

@article{steiner2010universal,
  title={Universal enhancement of the optical readout fidelity of single electron spins at nitrogen-vacancy centers in diamond},
  author={Steiner, M and Neumann, P and Beck, J and Jelezko, F and Wrachtrup, J},
  journal={Phys. Rev. B},
  volume={81},
  number={3},
  pages={035205},
  year={2010},
  publisher={APS}
}

@article{poulsen2022optimal,
  title={Optimal control of a nitrogen-vacancy spin ensemble in diamond for sensing in the pulsed domain},
  author={Poulsen, Andreas FL and Clement, Joshua D and Webb, James L and Jensen, Rasmus H and Troise, Luca and Berg-S{\o}rensen, Kirstine and Huck, Alexander and Andersen, Ulrik Lund},
  journal={Phys. Rev. B},
  volume={106},
  number={1},
  pages={014202},
  year={2022},
  publisher={APS}
}

@article{hermann2024extending,
  title={Extending radiowave frequency detection range with dressed states of solid-state spin ensembles},
  author={Hermann, Jens C and Rizzato, Roberto and Bruckmaier, Fleming and Allert, Robin D and Blank, Aharon and Bucher, Dominik B},
  journal={npj Quantum Inf.},
  volume={10},
  number={1},
  pages={103},
  year={2024},
  publisher={Nature Publishing Group UK London}
}

@article{delord2020spin,
  title={Spin-cooling of the motion of a trapped diamond},
  author={Delord, Tom and Huillery, P and Nicolas, L and H{\'e}tet, G},
  journal={Nature},
  volume={580},
  number={7801},
  pages={56--59},
  year={2020},
  publisher={Nature Publishing Group UK London}
}

@article{orphal2025coherent,
  title={Coherent microwave, optical, and mechanical quantum control of spin qubits in diamond},
  author={Orphal-Kobin, Laura and Torun, Cem G{\"u}ney and Bopp, Julian M and Pieplow, Gregor and Schr{\"o}der, Tim},
  journal={Adv. Quantum Technol.},
  volume={8},
  number={2},
  pages={2300432},
  year={2025},
  publisher={Wiley Online Library}
}

@article{pfender2017nonvolatile,
  title={Nonvolatile nuclear spin memory enables sensor-unlimited nanoscale spectroscopy of small spin clusters},
  author={Pfender, Matthias and Aslam, Nabeel and Sumiya, Hitoshi and Onoda, Shinobu and Neumann, Philipp and Isoya, Junichi and Meriles, Carlos A and Wrachtrup, J{\"o}rg},
  journal={Nat. Commun.},
  volume={8},
  number={1},
  pages={834},
  year={2017},
  publisher={Nature Publishing Group}
}

@article{dolde2014high,
  title={High-fidelity spin entanglement using optimal control},
  author={Dolde, Florian and Bergholm, Ville and Wang, Ya and Jakobi, Ingmar and Naydenov, Boris and Pezzagna, S{\'e}bastien and Meijer, Jan and Jelezko, Fedor and Neumann, Philipp and Schulte-Herbr{\"u}ggen, Thomas and others},
  journal={Nat. Commun.},
  volume={5},
  number={1},
  pages={3371},
  year={2014},
  publisher={Nature Publishing Group UK London}
}

@article{barry2020sensitivity,
  title={Sensitivity optimization for NV-diamond magnetometry},
  author={Barry, John F and Schloss, Jennifer M and Bauch, Erik and Turner, Matthew J and Hart, Connor A and Pham, Linh M and Walsworth, Ronald L},
  journal={Rev. of Mod. Phys.},
  volume={92},
  number={1},
  pages={015004},
  year={2020},
  publisher={APS}
}

@article{cai2012robust,
  title={Robust dynamical decoupling with concatenated continuous driving},
  author={Cai, JM and Naydenov, Boris and Pfeiffer, Rainer and McGuinness, Liam P and Jahnke, Kay D and Jelezko, Fedor and Plenio, Martin B and Retzker, Alex},
  journal={New J. Phys.},
  volume={14},
  number={11},
  pages={113023},
  year={2012},
  publisher={IOP Publishing}
}

@article{xu2012coherence,
  title={Coherence-protected quantum gate by continuous dynamical decoupling in diamond},
  author={Xu, Xiangkun and Wang, Zixiang and Duan, Changkui and Huang, Pu and Wang, Pengfei and Wang, Ya and Xu, Nanyang and Kong, Xi and Shi, Fazhan and Rong, Xing and others},
  journal={Phys. Rev. Lett.},
  volume={109},
  number={7},
  pages={070502},
  year={2012},
  publisher={APS}
}

@article{macquarrie2015continuous,
  title={Continuous dynamical decoupling of a single diamond nitrogen-vacancy center spin with a mechanical resonator},
  author={MacQuarrie, ER and Gosavi, TA and Bhave, SA and Fuchs, GD},
  journal={Phys. Rev. B},
  volume={92},
  number={22},
  pages={224419},
  year={2015},
  publisher={APS}
}

@article{gao2026,
  title = {Dressed-State Hamiltonian Engineering in a Strongly Interacting Solid-State Spin Ensemble},
  author = {Gao, Haoyang and Leitao, Nathaniel T. and Dandavate, Siddharth and Wyatt, Lillian B. Hughes and Put, Piotr and Mammen, Mathew and Martin, Leigh S. and Park, Hongkun and Jayich, Ania C. Bleszynski and Lukin, Mikhail D.},
  journal = {Phys. Rev. Lett.},
  year = {2026},
  month = {may},
  volume = {136},
  number = {20},
  pages = {200802},
  publisher = {American Physical Society (APS)},
  doi = {10.1103/d1kx-hy93},
  pmid = {42251591}
}

@article{salhov2024,
  title = {Protecting Quantum Information via Destructive Interference of Correlated Noise},
  author = {Salhov, Alon and Cao, Qingyun and Cai, Jianming and Retzker, Alex and Jelezko, Fedor and Genov, Genko},
  journal = {Phys. Rev. Lett.},
  year = {2024},
  month = {may},
  volume = {132},
  number = {22},
  pages = {223601},
  publisher = {American Physical Society (APS)},
  doi = {10.1103/physrevlett.132.223601},
  pmid = {38877916},
  url = {http://link.aps.org/pdf/10.1103/PhysRevLett.132.223601}
}

@article{aiello2013,
  title = {Composite-pulse magnetometry with a solid-state quantum sensor},
  author = {Aiello, Clarice D. and Hirose, Masashi and Cappellaro, Paola},
  journal = {Nat. Commun.},
  year = {2013},
  month = {jan},
  volume = {4},
  number = {1},
  pages = {1419},
  publisher = {Springer Nature},
  doi = {10.1038/ncomms2375},
  pmid = {23361010},
  url = {https://www.nature.com/articles/ncomms2375.pdf}
}

@article{wang2022,
  title = {Picotesla magnetometry of microwave fields with diamond sensors},
  author = {Wang, Zhecheng and Kong, Fei and Zhao, Pengju and Huang, Zhehua and Yu, Pei and Wang, Ya and Shi, Fazhan and Du, Jiangfeng},
  journal = {Sci. Adv.},
  year = {2022},
  month = {aug},
  volume = {8},
  number = {32},
  pages = {eabq8158},
  publisher = {American Association for the Advancement of Science (AAAS)},
  doi = {10.1126/sciadv.abq8158},
  pmid = {35947671},
  url = {https://doi.org/10.1126/sciadv.abq8158}
}

@article{wang2021,
  title = {Nanoscale Vector AC Magnetometry with a Single Nitrogen-Vacancy Center in Diamond},
  author = {Wang, Guoqing and Liu, Yi-Xiang and Zhu, Yuan and Cappellaro, Paola},
  journal = {Nano Lett.},
  year = {2021},
  month = {jun},
  volume = {21},
  number = {12},
  pages = {5143-5150},
  publisher = {American Chemical Society (ACS)},
  doi = {10.1021/acs.nanolett.1c01165},
  pmid = {34086471},
  url = {https://arxiv.org/pdf/2103.12044}
}

@article{liu2026,
  title = {Strain-Engineered Nanoscale Spin Polarization Reversal in Diamond Nitrogen-Vacancy Centers},
  author = {Liu, Zhixian and Sun, Jiahao and Xu, Ganyu and Yang, Bo and Guo, Yuhang and Wang, Yu and Xin, Cunliang and Zuo, Hongfang and Wang, Mengqi and Wang, Ya},
  journal = {Phys. Rev. Lett.},
  year = {2026},
  month = {apr},
  volume = {136},
  number = {14},
  pages = {143001},
  publisher = {American Physical Society (APS)},
  doi = {10.1103/by4s-xbbn},
  pmid = {42033705}
}

@article{delord2025,
  title = {Probing Electric-Dipole-Enabled Transitions in the Excited State of the Nitrogen-Vacancy Center in Diamond},
  author = {Delord, Tom and Monge, Richard and L\'{o}pez-Morales, Gabriel I. and Bach, Olaf and Dreyer, Cyrus E. and Flick, Johannes and Meriles, Carlos A.},
  journal = {Phys. Rev. Lett.},
  year = {2025},
  month = {nov},
  volume = {135},
  number = {22},
  pages = {226401},
  publisher = {American Physical Society (APS)},
  doi = {10.1103/hp9c-rh23},
  pmid = {41385702}
}

@article{rathi2026,
  title = {Engineering Nanodiamonds for Quantum Sensing: Material Constraints at the Nanoscale},
  author = {Rathi, Ashutosh and Oshimi, Keisuke and Sasaki, Kento and Kobayashi, Kensuke and Shikano, Yutaka and Benson, Oliver and Schr\"{o}der, Tim and Chang, Shery L. Y. and Fujiwara, Masazumi},
  journal = {ACS Nano},
  year = {2026},
  month = {jun},
  volume = {20},
  number = {24},
  pages = {17143-17162},
  publisher = {American Chemical Society (ACS)},
  doi = {10.1021/acsnano.6c03795},
  pmid = {42261724},
  url = {https://pubs.acs.org/doi/pdf/10.1021/acsnano.6c03795?ref=article\_openPDF}
}

@article{udvarhelyi2018,
  title = {Spin-strain interaction in nitrogen-vacancy centers in diamond},
  author = {Udvarhelyi, P\'{e}ter and Shkolnikov, V. O. and Gali, Adam and Burkard, Guido and P\'{a}lyi, Andr\'{a}s},
  journal = {Phys. Rev. B},
  year = {2018},
  month = {aug},
  volume = {98},
  number = {7},
  pages = {075201},
  publisher = {American Physical Society (APS)},
  doi = {10.1103/physrevb.98.075201},
  url = {https://arxiv.org/pdf/1712.02684}
}

@article{louzon2025,
  title = {Robust Noise Suppression and Quantum Sensing by Continuous Phased Dynamical Decoupling},
  author = {Louzon, Daniel and Genov, Genko T. and Staudenmaier, Nicolas and Frank, Florian and Lang, Johannes and Markham, Matthew L. and Retzker, Alex and Jelezko, Fedor},
  journal = {Phys. Rev. Lett.},
  year = {2025},
  month = {mar},
  volume = {134},
  number = {12},
  pages = {120802},
  publisher = {American Physical Society (APS)},
  doi = {10.1103/physrevlett.134.120802},
  pmid = {40215502},
  url = {http://link.aps.org/pdf/10.1103/PhysRevLett.134.120802}
}

@article{stark2017,
  title = {Narrow-bandwidth sensing of high-frequency fields with continuous dynamical decoupling},
  author = {Stark, Alexander and Aharon, Nati and Unden, Thomas and Louzon, Daniel and Huck, Alexander and Retzker, Alex and Andersen, Ulrik L. and Jelezko, Fedor},
  journal = {Nat. Commun.},
  year = {2017},
  month = {oct},
  volume = {8},
  number = {1},
  pages = {1105},
  publisher = {Springer Nature},
  doi = {10.1038/s41467-017-01159-2},
  pmid = {29051547},
  url = {https://www.nature.com/articles/s41467-017-01159-2.pdf}
}

@article{kollarics2024,
  title = {Terahertz emission from diamond nitrogen-vacancy centers},
  author = {Kollarics, S\'{a}ndor and M\'{a}rkus, Bence G\'{a}bor and Kucsera, Robin and Thiering, Gerg\H{o} and Gali, \'{A}d\'{a}m and N\'{e}meth, Gergely and Kamar\'{a}s, Katalin and Forr\'{o}, L\'{a}szl\'{o} and Simon, Ferenc},
  journal = {Sci. Adv.},
  year = {2024},
  month = {may},
  volume = {10},
  number = {22},
  pages = {eadn0616},
  publisher = {American Association for the Advancement of Science (AAAS)},
  doi = {10.1126/sciadv.adn0616},
  pmid = {38809991},
  url = {https://www.science.org/doi/pdf/10.1126/sciadv.adn0616?download=true}
}

@article{apra2025,
  title = {Effects of Thermal Oxidation and Proton Irradiation on Optically Detected Magnetic Resonance Sensitivity in Sub-100 nm Nanodiamonds},
  author = {Apr\`{a}, Pietro and Zanelli, Gabriele and Losero, Elena and Amine, Nour-Hanne and Andrini, Greta and Barozzi, Mario and Bernardi, Ettore and Britel, Adam and Canteri, Roberto and Degiovanni, Ivo Pietro and Mino, Lorenzo and Moreva, Ekaterina and Olivero, Paolo and Redolfi, Elisa and Stella, Claudia and Sturari, Sofia and Traina, Paolo and Varzi, Veronica and Genovese, Marco and Picollo, Federico},
  journal = {ACS Appl. Mater. Interf.},
  year = {2025},
  month = {mar},
  volume = {17},
  number = {14},
  pages = {21589-21600},
  publisher = {American Chemical Society (ACS)},
  doi = {10.1021/acsami.4c08780},
  pmid = {40159101},
  url = {https://pubs.acs.org/doi/pdf/10.1021/acsami.4c08780?ref=article\_openPDF}
}

@article{ovartchaiyapong2014dynamic,
  title={Dynamic strain-mediated coupling of a single diamond spin to a mechanical resonator},
  author={Ovartchaiyapong, Preeti and Lee, Kenneth W and Myers, Bryan A and Jayich, Ania C Bleszynski},
  journal={Nature communications},
  volume={5},
  number={1},
  pages={4429},
  year={2014},
  publisher={Nature Publishing Group UK London}
}

@article{meesala2016enhanced,
  title={Enhanced strain coupling of nitrogen-vacancy spins to nanoscale diamond cantilevers},
  author={Meesala, Srujan and Sohn, Young-Ik and Atikian, Haig A and Kim, Samuel and Burek, Michael J and Choy, Jennifer T and Lon{\v{c}}ar, Marko},
  journal={Phys. Rev. Appl.},
  volume={5},
  number={3},
  pages={034010},
  year={2016},
  publisher={APS}
}

@article{fuchs2008excited,
  title={Excited-state spectroscopy using single spin manipulation in diamond},
  author={Fuchs, GD and Dobrovitski, VV and Hanson, R and Batra, A and Weis, CD and Schenkel, T and Awschalom, DD},
  journal={Phys. Rev. Lett.},
  volume={101},
  number={11},
  pages={117601},
  year={2008},
  publisher={APS}
}

@article{ten2013stochastic,
  title={Stochastic determination of effective Hamiltonian for the full configuration interaction solution of quasi-degenerate electronic states},
  author={Ten-no, Seiichiro},
  journal={J. Chem. Phys.},
  volume={138},
  number={16},
  year={2013},
  publisher={AIP Publishing}
}

@article{pavlyukh2015single,
  title={Single-or double-electron emission within the Keldysh nonequilibrium Green's function and Feshbach projection operator techniques},
  author={Pavlyukh, Y and Sch{\"u}ler, M and Berakdar, J},
  journal={Phys. Rev. B},
  volume={91},
  number={15},
  pages={155116},
  year={2015},
  publisher={APS}
}

@article{shamshutdinova2008feshbach,
  title={Feshbach projection-operator formalism applied to resonance scattering on Bargmann-type potentials},
  author={Shamshutdinova, Varvara V and Pichugin, Konstantin N and Rotter, Ingrid and Samsonov, Boris F},
  journal={Phys. Rev. A},
  volume={78},
  number={6},
  pages={062712},
  year={2008},
  publisher={APS}
}

@article{christopher2006efficient,
  title={Efficient partitioning technique for computing the dynamics of intramolecular processes: Radiationless transitions in pyrazine},
  author={Christopher, PS and Shapiro, Moshe and Brumer, Paul},
  journal={J. Chem. Phys.},
  volume={124},
  number={18},
  year={2006},
  publisher={AIP Publishing}
}

@article{zhang1990photodissociation,
  title={Photodissociation and continuum resonance Raman cross sections and general Franck--Condon intensities from S-matrix Kohn scattering calculations with application to the photoelectron spectrum of H2F-+ h $\nu$→ H2+ F, HF+ H+ e-},
  author={Zhang, John ZH and Miller, William H},
  journal={J. Chem. Phys.},
  volume={92},
  number={3},
  pages={1811--1818},
  year={1990},
  publisher={American Institute of Physics}
}

@article{jin2011partitioning,
  title={Partitioning technique for discrete quantum systems},
  author={Jin, L and Song, Z},
  journal={Phys. Rev. A},
  volume={83},
  number={6},
  pages={062118},
  year={2011},
  publisher={APS}
}

@article{kolbl2019initialization,
  title={Initialization of single spin dressed states using shortcuts to adiabaticity},
  author={K{\"o}lbl, Johannes and Barfuss, Arne and Kasperczyk, MS and Thiel, Lucas and Clerk, AA and Ribeiro, Hugo and Maletinsky, Patrick},
  journal={Phys. Rev. Lett.},
  volume={122},
  number={9},
  pages={090502},
  year={2019},
  publisher={APS}
}

@article{rabl2009strong,
  title={Strong magnetic coupling between an electronic spin qubit and a mechanical resonator},
  author={Rabl, Peter and Cappellaro, P and Dutt, MV Gurudev and Jiang, Liang and Maze, JR and Lukin, Mikhail D},
  journal={Phys. Rev. B},
  volume={79},
  number={4},
  pages={041302},
  year={2009},
  publisher={APS}
}

@article{belthangady2013dressed,
  title={Dressed-state resonant coupling between bright and dark spins in diamond},
  author={Belthangady, Chinmay and Bar-Gill, Nir and Pham, Linh My and Arai, Keigo and Le Sage, David and Cappellaro, Paola and Walsworth, Ronald Lee},
  journal={Phys. Rev. Lett.},
  volume={110},
  number={15},
  pages={157601},
  year={2013},
  publisher={APS}
}

@article{timoney2011quantum,
  title={Quantum gates and memory using microwave-dressed states},
  author={Timoney, N and Baumgart, I and Johanning, M and Var{\'o}n, AF and Plenio, Martin B and Retzker, A and Wunderlich, Ch},
  journal={Nature},
  volume={476},
  number={7359},
  pages={185--188},
  year={2011},
  publisher={Nature Publishing Group UK London}
}

@article{manson2006nitrogen,
  title={Nitrogen-vacancy center in diamond: Model of the electronic structure and associated dynamics},
  author={Manson, NB and Harrison, JP and Sellars, MJ},
  journal={Phys. Rev. B},
  volume={74},
  number={10},
  pages={104303},
  year={2006},
  publisher={APS}
}

@article{hodges2013timekeeping,
  title={Timekeeping with electron spin states in diamond},
  author={Hodges, Jonathan S and Yao, Norman Ying and Maclaurin, Dougal and Rastogi, C and Lukin, Mikhail D and Englund, D},
  journal={Phys. Rev. A},
  volume={87},
  number={3},
  pages={032118},
  year={2013},
  publisher={APS}
}

@article{happacher2022low,
  title={Low-temperature photophysics of single nitrogen-vacancy centers in diamond},
  author={Happacher, Jodok and Broadway, David A and Bocquel, Juanita and Reiser, Patrick and Jimen{\'e}z, Alejandro and Tschudin, M{\"a}rta A and Thiel, Lucas and Rohner, Dominik and Puigibert, Marcel li Grimau and Shields, Brendan and others},
  journal={Phys. Rev. Lett.},
  volume={128},
  number={17},
  pages={177401},
  year={2022},
  publisher={APS}
}

@article{goldman2015state,
  title={State-selective intersystem crossing in nitrogen-vacancy centers},
  author={Goldman, Michael Lurie and Doherty, MW and Sipahigil, Alp and Yao, Norman Ying and Bennett, SD and Manson, NB and Kubanek, Alexander and Lukin, Mikhail D},
  journal={Phys. Rev. B},
  volume={91},
  number={16},
  pages={165201},
  year={2015},
  publisher={APS}
}

@article{gupta2016efficient,
  title={Efficient signal processing for time-resolved fluorescence detection of nitrogen-vacancy spins in diamond},
  author={Gupta, Anchal and Hacquebard, Luke and Childress, Lilian},
  journal={J. Opt. Soc. Am. B},
  volume={33},
  number={3},
  pages={B28--B34},
  year={2016},
  publisher={Optical Society of America}
}

@article{chapman2012anomalous,
  title={Anomalous saturation effects due to optical spin depolarization in nitrogen-vacancy centers in diamond nanocrystals},
  author={Chapman, Robert and Plakhotnik, Taras},
  journal={Phys. Rev. B},
  volume={86},
  number={4},
  pages={045204},
  year={2012},
  publisher={APS}
}

@article{ernst2023modeling,
  title={Modeling temperature-dependent population dynamics in the excited state of the nitrogen-vacancy center in diamond},
  author={Ernst, Stefan and Scheidegger, Patrick J and Diesch, Simon and Degen, Christian L},
  journal={Phys. Rev. B},
  volume={108},
  number={8},
  pages={085203},
  year={2023},
  publisher={APS}
}

@article{lee2016strain,
  title={Strain coupling of a mechanical resonator to a single quantum emitter in diamond},
  author={Lee, Kenneth W and Lee, Donghun and Ovartchaiyapong, Preeti and Minguzzi, Joaquin and Maze, Jero R and Bleszynski Jayich, Ania C},
  journal={Phys. Rev. Appl.},
  volume={6},
  number={3},
  pages={034005},
  year={2016},
  publisher={APS}
}

@article{dolde2014nanoscale,
  title={Nanoscale detection of a single fundamental charge in ambient conditions using the NV- center in diamond},
  author={Dolde, Florian and Doherty, Marcus W and Michl, Julia and Jakobi, Ingmar and Naydenov, Boris and Pezzagna, Sebastien and Meijer, Jan and Neumann, Philipp and Jelezko, Fedor and Manson, Neil B and others},
  journal={Physical review letters},
  volume={112},
  number={9},
  pages={097603},
  year={2014},
  publisher={APS}
}

@article{acosta2012dynamic,
  title={Dynamic Stabilization of the Optical Resonances of Single Nitrogen-Vacancy Centers in Diamond},
  author={Acosta, VM and Santori, C and Faraon, A and Huang, Z and Fu, K-MC and Stacey, Alastair and Simpson, DA and Ganesan, K and Tomljenovic-Hanic, S and Greentree, AD and others},
  journal={Phys. Rev. Lett.},
  volume={108},
  number={20},
  pages={206401},
  year={2012},
  publisher={APS}
}

@article{poem2015broadband,
  title={Broadband noise-free optical quantum memory with neutral nitrogen-vacancy centers in diamond},
  author={Poem, E and Weinzetl, C and Klatzow, J and Kaczmarek, KT and Munns, JHD and Champion, TFM and Saunders, DJ and Nunn, J and Walmsley, IA},
  journal={Phys. Rev. B},
  volume={91},
  number={20},
  pages={205108},
  year={2015},
  publisher={APS}
}

@article{kuno2026concatenated,
  title={Concatenated continuous driving of silicon qubit by amplitude and phase modulation},
  author={Kuno, Takuma and Utsugi, Takeru and Ramsay, Andrew J and Mertig, Normann and Lee, Noriyuki and Yanagi, Itaru and Mine, Toshiyuki and Kusuno, Nobuhiro and Arimoto, Hideo and Beyne, Sofie and others},
  journal={Phys. Rev. B},
  volume={113},
  number={19},
  pages={195303},
  year={2026},
  publisher={APS}
}

\appendix
\section{Effective two-level reduction}\label{app:reduction}

We detail the L\"owdin-Feshbach reduction of the dressed Hamiltonian $H_{\mathrm{dr}}$ [Eq.~\eqref{eq:hdr}]. We define the frequency Hamiltonian $\mathcal H_{\mathrm{dr}}\equiv H_{\mathrm{dr}}/\hbar$ and partition the three-level space with $P=\ket{0}\bra{0}+\ket{\widetilde{-}}\bra{\widetilde{-}}$ onto the qubit sector and $Q=\ket{\widetilde+}\bra{\widetilde+}$ onto the spectator. In the ordered qubit basis $\{\ket0,\ket{\widetilde-}\}$, the corresponding blocks are
\begin{align}
P\mathcal H_{\mathrm{dr}}P=\begin{pmatrix}0&\Omega_-/2\\ \Omega_-/2&0\end{pmatrix},\qquad P\mathcal H_{\mathrm{dr}}Q=\begin{pmatrix}\Omega_+/2\\ 0\end{pmatrix},\qquad Q\mathcal H_{\mathrm{dr}}Q=\Delta ,
\end{align}
so that only $\ket0$ couples to the spectator. The exact effective frequency Hamiltonian $\mathcal H_{\mathrm{eff}}(\lambda)\equiv H_{\mathrm{eff}}(E)/\hbar$ at the qubit eigenfrequency $\lambda=E/\hbar$ is
\begin{align}
\mathcal H_{\mathrm{eff}}(\lambda)=P\mathcal H_{\mathrm{dr}}P+P\mathcal H_{\mathrm{dr}}Q\,\left(\lambda-Q\mathcal H_{\mathrm{dr}}Q\right)^{-1}\,Q\mathcal H_{\mathrm{dr}}P=\begin{pmatrix}\dfrac{\Omega_+^2/4}{\lambda-\Delta}&\dfrac{\Omega_-}{2}\\[6pt]
\dfrac{\Omega_-}{2}&0\end{pmatrix}.
\end{align}
In the far-detuned regime, $|\lambda|\ll\Delta$, we evaluate the resolvent at $\lambda\to0$. This gives $(\lambda-\Delta)^{-1}\simeq-1/\Delta$ and shifts $\ket0$ by $-\Omega_+^2/4\Delta$. With $\Delta_{\mathrm{eff}}\equiv-\Omega_+^2/4\Delta$, one obtains $\mathcal H_{\mathrm{eff}}=\bigl(\begin{smallmatrix}\Delta_{\mathrm{eff}}&\Omega_-/2\\ \Omega_-/2&0\end{smallmatrix}\bigr)$. Setting $\sigma_z=\ket0\bra0-\ket{\widetilde-}\bra{\widetilde-}$, this Hamiltonian can be written as $\mathcal H_{\mathrm{eff}}=(\Delta_{\mathrm{eff}}/2)\mathbb{1}+\tfrac12(\Delta_{\mathrm{eff}}\sigma_z+\Omega_-\sigma_x)$. Discarding the common shift $\Delta_{\mathrm{eff}}/2$ gives the centered two-level Hamiltonian. Its eigenvalues are split by $\Omega_D=\sqrt{\Delta_{\mathrm{eff}}^2+\Omega_-^2}$, and the mixing angle satisfies $\tan\theta=\Omega_-/\Delta_{\mathrm{eff}}$, with the branch chosen continuously such that $\theta\to\pi/2$ as $g_\perp\to0$. Evaluating the spectator denominator at the qubit centroid gives the simpler far-detuned estimate $|\Omega_+|/(2\Delta)$. This centroid-based estimate differs from the branch-specific $\varepsilon$ in Eq. \eqref{eq:epsilon} and is not used as the validity boundary.

The longitudinal probe couples the dressed states through $S_z$. In the bare basis $S_z=\mathrm{diag}(1,0,-1)$; projected onto the qubit sector, one has $\bra{0}S_z\ket{0}=0$, $\bra{\widetilde-}S_z\ket{\widetilde-}=-\cos2\psi$, and $\bra{0}S_z\ket{\widetilde-}=0$. Evaluating the matrix element between $\ket{d_\pm}$ gives $m_q=\left|\bra{d_+}S_z\ket{d_-}\right|=\tfrac12\cos2\psi\,\sin\theta$.

\section{Exact validity criteria}\label{app:validity}

We now derive the exact criteria used to validate the approximate diagram in Sec. \ref{sec:validity}. We restrict the discussion to $B_z\geq0$ and $g_\perp\geq0$, as in Fig. \ref{fig:phase}, and retain the physical parameters already defined in the main text. The full dressed Hamiltonian is
\begin{align}
 \frac{H_{\mathrm{dr}}}{\hbar}
 =\begin{pmatrix}
 \Delta&\Omega_+/2&0\\
 \Omega_+/2&0&\Omega_-/2\\
 0&\Omega_-/2&0
 \end{pmatrix}.
 \label{eq:exact_h}
\end{align}
The ordered exact eigenfrequencies $\lambda_1<\lambda_2<\lambda_3$ are the three real roots of
\begin{align}
 0={}&\det\!\left(\lambda\mathbb{1}-H_{\mathrm{dr}}/\hbar\right)\nonumber\\
 ={}&\lambda^3-\Delta\lambda^2
 -\frac{\Omega_+^2+\Omega_-^2}{4}\lambda
 +\frac{\Delta\Omega_-^2}{4}\nonumber\\
 ={}&\lambda^3-\Delta\lambda^2-\frac{\Omega^2}{4}\lambda
 +\frac{\Delta\Omega_-^2}{4}.
 \label{eq:exact_cubic}
\end{align}
This cubic equation provides the exact eigenfrequencies without invoking the spectator-state elimination.

For an eigenfrequency $\lambda_j$, the corresponding eigenvector can be written, before normalization, as
\begin{align}
 \ket{\lambda_j}\propto
 \frac{\Omega_+/2}{\lambda_j-\Delta}\ket{\widetilde+}
 +\ket0+\frac{\Omega_-/2}{\lambda_j}\ket{\widetilde-}.
 \label{eq:exact_eigenvector}
\end{align}
The spectator weight of this state is therefore
\begin{align}
 w_j&\equiv\left|\braket{\widetilde+|\lambda_j}\right|^2\nonumber\\
 &=\frac{\Omega_+^2/[4(\lambda_j-\Delta)^2]}
 {\Omega_+^2/[4(\lambda_j-\Delta)^2]+1+\Omega_-^2/(4\lambda_j^2)}.
 \label{eq:spectator_weight}
\end{align}
At $\Omega_+=0$, Eq. \eqref{eq:spectator_weight} is understood by continuity. In this limit, one exact eigenstate is the uncoupled spectator with unit spectator weight.

For the numerical benchmark, we identify the state with the largest $w_j$ as the spectator-like branch and order the remaining two qubit-character branches as $\lambda_{\mathrm{q,l}}<\lambda_{\mathrm{q,u}}$. The exact dressed splitting is
\begin{align}
    \Omega_D^{(3)}=\lambda_{\mathrm{q,u}}-\lambda_{\mathrm{q,l}},
    \label{eq:OmegaD_full}
\end{align}
whereas Eq. \eqref{eq:OmegaD_reduced} defines the reduced result $\Omega_D^{(2)}$. To extract a comparable spin-locking angle from the full model, we remove the spectator component of the upper qubit-character state and normalize its projection onto the nominal qubit subspace,
\begin{align}
    \frac{P\ket{\lambda_{\mathrm{q,u}}}}
    {\sqrt{\bra{\lambda_{\mathrm{q,u}}}P\ket{\lambda_{\mathrm{q,u}}}}}
    =\cos\!\left(\frac{\theta^{(3)}}{2}\right)\ket0
    +\sin\!\left(\frac{\theta^{(3)}}{2}\right)\ket{\widetilde-}.
    \label{eq:full_projected_angle}
\end{align}
The overall phase is chosen continuously along each parameter sweep. We use $\theta^{(2)}=\theta$ for the reduced model and define $\Delta\theta^{(n)}=\theta^{(n)}-\pi/2$ for $n=2,3$.

In the parameter range of Fig. \ref{fig:phase}, the lower branch $\lambda_1$ remains predominantly qubit-like, while the spectator character is exchanged between $\lambda_2$ and $\lambda_3$. For $w_3>w_2$, the highest branch $\lambda_3$ is spectator-like, and the two qubit-character branches are $\lambda_1$ and $\lambda_2$. For $w_2>w_3$, the middle branch $\lambda_2$ is spectator-like and is interleaved between the two qubit-character branches. The exact character-exchange boundary is therefore
\begin{align}
 \boxed{w_2(B_z,g_\perp)=w_3(B_z,g_\perp).}
 \label{eq:exact_primary}
\end{align}
The full eigenvalues remain continuous at this boundary. Equation \eqref{eq:exact_primary} identifies the point at which the dominant spectator assignment changes between two exact branches.

On the $w_3>w_2$ side, the exact ratio of the spectator amplitude to the qubit-subspace amplitude in an exact qubit branch is $\sqrt{w_j/(1-w_j)}$. Retaining the practical amplitude threshold $0.05$, the three regions are defined without further approximation as
\begin{align}
 \text{region I:}\quad&w_3>w_2,\qquad
 \max_{j=1,2}\sqrt{\frac{w_j}{1-w_j}}<0.05,\nonumber\\
 \text{region II:}\quad&w_3>w_2,\qquad
 \max_{j=1,2}\sqrt{\frac{w_j}{1-w_j}}\geq0.05,\nonumber\\
 \text{region III:}\quad&w_2>w_3.
 \label{eq:exact_regions}
\end{align}
The amplitude threshold $0.05$ corresponds to an exact spectator probability of $0.05^2/(1+0.05^2)\simeq2.49\times10^{-3}$. Unlike the character boundary in Eq. \eqref{eq:exact_primary}, the I--II boundary is tolerance dependent because the error of an effective Hamiltonian changes continuously.

The relation between the exact and approximate boundaries follows directly by using the nominal dressed states in Eq. \eqref{eq:nominal_dressed}. The superscript $(0)$ indicates that the coupling to $\ket{\widetilde+}$ is omitted while the strain dependence of $\ket{\widetilde-}$ and $\Omega_-$ is retained. In the ordered basis $\{\ket{\widetilde+},\ket{d_+^{(0)}},\ket{d_-^{(0)}}\}$, Eq. \eqref{eq:exact_h} becomes
\begin{align}
 \frac{H_{\mathrm{dr}}}{\hbar}=\begin{pmatrix}
 \Delta&\Omega_+/(2\sqrt2)&\Omega_+/(2\sqrt2)\\
 \Omega_+/(2\sqrt2)&\Omega_-/2&0\\
 \Omega_+/(2\sqrt2)&0&-\Omega_-/2
 \end{pmatrix}.
 \label{eq:diabatic_basis}
\end{align}
Neglecting the coupling-induced displacement from the lower state $\ket{d_-^{(0)}}$ gives the diabatic character boundary $\Delta=\Omega_-/2$, which is Eq. \eqref{eq:guideline_primary}. The coupling matrix element and the unperturbed spectral detuning relevant to the nominal upper dressed state are, respectively,
\begin{align}
 \left|\bra{\widetilde+}\frac{H_{\mathrm{dr}}}{\hbar}\ket{d_+^{(0)}}\right|
 &=\frac{|\Omega_+|}{2\sqrt2},\qquad
 \Delta-\frac{\Omega_-}{2}.
 \label{eq:upper_coupling_detuning}
\end{align}
Accordingly, on the $\Delta>\Omega_-/2$ side and in the weak-mixing limit $\varepsilon\ll1$, first-order nondegenerate perturbation theory gives the following normalized approximation to the upper qubit branch:
\begin{align}
 \ket{\lambda_2}\simeq
 \frac{
 \displaystyle \ket{d_+^{(0)}}-
 \frac{\Omega_+/(2\sqrt2)}{\Delta-\Omega_-/2}\ket{\widetilde+}
 }{
 \displaystyle \sqrt{1+\frac{\Omega_+^2/8}{\left(\Delta-\Omega_-/2\right)^2}}
 }.
 \label{eq:upper_perturbative_state}
\end{align}
The magnitude of the spectator-to-qubit amplitude ratio in Eq. \eqref{eq:upper_perturbative_state} is therefore
\begin{align}
 \frac{|\Omega_+|/(2\sqrt2)}{\Delta-\Omega_-/2}=\varepsilon,
 \label{eq:epsilon_perturbative}
\end{align}
and the corresponding spectator probability is
\begin{align}
 \frac{\varepsilon^2}{1+\varepsilon^2}\simeq\varepsilon^2.
 \label{eq:epsilon_probability}
\end{align}
For the nominal lower dressed state, the analogous first-order amplitude ratio is smaller,
\begin{align}
 \frac{|\Omega_+|/(2\sqrt2)}{\Delta+\Omega_-/2}<\varepsilon,
 \label{eq:lower_admixture}
\end{align}
so the upper dressed branch controls the approximate I--II boundary. In particular, the chosen value $\varepsilon=0.05$ corresponds to a spectator probability $0.05^2/(1+0.05^2)\simeq2.49\times10^{-3}$. Thus, Eqs. \eqref{eq:guideline_primary} and \eqref{eq:guideline_secondary} are controlled approximations to Eqs. \eqref{eq:exact_primary} and \eqref{eq:exact_regions}, respectively. For $B_z=1~\mathrm{G}$ and $\Omega=16~\mathrm{MHz}$, the exact condition $w_2=w_3$ gives $g_\perp\simeq2.32~\mathrm{MHz}$, while the diabatic approximation gives $2.48~\mathrm{MHz}$.

\section{Dark-phase ESR dynamics}\label{app:me}

The pulsed ESR sequence consists of state preparation, dark evolution, and analysis, followed by optical readout. The optical initialization and preparation operations are not simulated explicitly. Instead, the calculation starts from the upper qubit-character branch, which gives $\rho_{++}(0)=1$ in the reduced model. During the dark interval, the laser is switched off, and a weak longitudinal probe with Rabi frequency $\Omega_{\mathrm{R}}$ and frequency $\omega_{\mathrm{p}}$ is applied for a time $\tau$. The analysis mapping and subsequent optical dynamics are described in Appendix \ref{app:obe}.

In the frame rotating at $\omega_{\mathrm{p}}$, and within the rotating-wave approximation, the effective two-level Hamiltonian is
\begin{align}
\frac{H_p}{\hbar}=\frac12\!\left[(\Omega_D^{(2)}-\omega_{\mathrm{p}})\,\sigma_z +\Omega_{\mathrm{R}}m_q\,\sigma_x\right],
\label{eq:reduced_probe_hamiltonian}
\end{align}
where $\sigma_z=\ket{d_+}\bra{d_+}-\ket{d_-}\bra{d_-}$ and the effective probe Rabi frequency is $\Omega_{\mathrm{R}}m_q$. We supplement the von Neumann equation with pure dephasing of the coherence at a rate $\gamma_2=1/T_{2\rho}$,
\begin{align}
\dot\rho=-\frac{i}{\hbar}[H_p,\rho]-\gamma_2\!\left(\rho-\mathrm{diag}\,\rho\right),
\end{align}
where the diagonal operation is taken in the $\{\ket{d_+},\ket{d_-}\}$ basis. Writing $\rho=\bigl(\begin{smallmatrix}\rho_{++}&u+iv\\ u-iv&\rho_{--}\end{smallmatrix}\bigr)$ gives
\begin{align}
\dot\rho_{++}&=-\Omega_{\mathrm{R}}m_q\,v, &
\dot\rho_{--}&=+\Omega_{\mathrm{R}}m_q\,v,\nonumber\\
\dot u&=(\Omega_D^{(2)}-\omega_{\mathrm{p}})v-\gamma_2 u, &
\dot v&=-(\Omega_D^{(2)}-\omega_{\mathrm{p}})u
+\tfrac12\Omega_{\mathrm{R}}m_q(\rho_{++}-\rho_{--})-\gamma_2 v.
\end{align}
Starting from $\rho_{++}(0)=1$ and integrating for a time $\tau$, the population $\rho_{++}(\tau)$ develops a dip at $\omega_{\mathrm{p}}=\Omega_D^{(2)}$. In the absence of dephasing, it is $\rho_{++}(\tau)=1-\left[(\Omega_{\mathrm{R}}m_q)^2/\Omega_g^2\right]\sin^2(\Omega_g\tau/2)$ with $\Omega_g=\sqrt{(\Omega_D^{(2)}-\omega_{\mathrm{p}})^2+(\Omega_{\mathrm{R}}m_q)^2}$. For the weak probe used here, the linewidth and sideband spacing are governed primarily by $1/\tau$.

For the full calculation, all three exact branches are retained, and the density matrix is evolved under
\begin{align}
    \frac{H_{\mathrm{p}}^{(3)}(t)}{\hbar}
    =\frac{H_{\mathrm{dr}}}{\hbar}
    +\Omega_{\mathrm{R}}\cos\!\left(\omega_{\mathrm{p}}t\right)S_z,
    \label{eq:full_probe_hamiltonian}
\end{align}
where the complete longitudinal spin operator in the ordered strain basis $\{\ket{\widetilde+},\ket0,\ket{\widetilde-}\}$ is
\begin{align}
    S_z=\begin{pmatrix}
    \cos2\psi&0&-\sin2\psi\\
    0&0&0\\
    -\sin2\psi&0&-\cos2\psi
    \end{pmatrix}.
    \label{eq:full_Sz}
\end{align}
The off-diagonal terms proportional to $\sin2\psi$ account for probe-induced transitions between the two strain eigenstates. We transform Eq. \eqref{eq:full_probe_hamiltonian} to the exact eigenbasis of $H_{\mathrm{dr}}$ and integrate
\begin{align}
    \dot\rho^{(3)}=-\frac{i}{\hbar}\left[H_{\mathrm{p}}^{(3)}(t),\rho^{(3)}\right]
    -\gamma_2\left(\rho^{(3)}-\operatorname{diag}\rho^{(3)}\right),
    \label{eq:full_master}
\end{align}
where the diagonal operation is taken in the exact eigenbasis. The upper qubit-character branch is used as the initial state. This calculation retains both the dressed-qubit transition and the additional transitions involving the spectator-like branch.

\section{Optical rate-equation readout}\label{app:obe}

The final dark-phase populations are converted into bare-spin populations by a frequency-independent projective analysis mapping\cite{kolbl2019initialization,rabl2009strong,xu2012coherence,belthangady2013dressed,timoney2011quantum}. In the reduced model, the upper and lower dressed populations are mapped to the $m_s=0$ and $m_s=-1$ ground states, respectively. In the full model, the populations of the upper qubit-character, lower qubit-character, and spectator-like branches seed the $m_s=0$, $m_s=-1$, and $m_s=+1$ ground states, respectively. This ideal mapping isolates the transitions generated during the dark interval without specifying a particular preparation or analysis-pulse waveform.

We model the subsequent spin-dependent PL using the standard NV rate-equation description\cite{manson2006nitrogen,hodges2013timekeeping,happacher2022low,goldman2015state,gupta2016efficient,liu2026,chapman2012anomalous,ernst2023modeling}. The model contains ten levels: the ground populations $g_{m_s}$, the virtual populations $v_{m_s}$, the excited populations $e_{m_s}$ ($m_s=0,\pm1$), and the effective shelving-singlet population $\rho_s$. Optical pumping (rate $W$), phonon-mediated relaxation (rate $\Gamma$), and radiative decay (rates $\gamma_{m_s}$) are spin conserving. Intersystem crossing (ISC) to the singlet ($k_{m_s}$) and the singlet returning to the ground states ($\kappa_{m_s}$) are spin selective, with $k_{\pm1}>k_0$ and $\kappa_0>\kappa_{\pm1}$. The populations evolve as
\begin{align}
\dot g_{m_s}&=-Wg_{m_s}+\gamma_{m_s}e_{m_s}+\kappa_{m_s}\rho_s,\nonumber\\
\dot v_{m_s}&=Wg_{m_s}-\Gamma v_{m_s},\nonumber\\
\dot e_{m_s}&=\Gamma v_{m_s}-(\gamma_{m_s}+k_{m_s})e_{m_s},\nonumber\\
\dot\rho_s&=\sum_{m_s}k_{m_s}e_{m_s}
-\left(\sum_{m_s}\kappa_{m_s}\right)\rho_s.
\end{align}
The populations are normalized such that $\sum_{m_s}(g_{m_s}+v_{m_s}+e_{m_s})+\rho_s=1$. The mapped bare-spin populations initialize $g_{m_s}$, with all optical and singlet populations initially set to zero. We then integrate the rate equations under continuous illumination. The instantaneous photon-emission rate and the time-integrated PL signal are
\begin{align}
    r_{\mathrm{PL}}(t)&=\sum_{m_s}\gamma_{m_s}e_{m_s}(t),\qquad
    I_{\mathrm{PL}}=\int_{0}^{T_{\mathrm{read}}}r_{\mathrm{PL}}(t)\,dt,
    \label{eq:integrated_PL}
\end{align}
where $\gamma_{m_s}$ are the spin-preserving radiative decay rates. Because the spin-selective ISC depletes $m_s=\pm1$ into the singlet, $m_s=0$ fluoresces more brightly than $m_s=\pm1$. The optical parameters are symmetric under $m_s=+1\leftrightarrow-1$, so the two dark channels have the same photon yield. The optical cycle therefore transduces the dark-phase populations without introducing an additional resonance-frequency dependence. Each calculated spectrum is divided by its own maximum PL value. Accordingly, the spectra are used to determine resonance positions and spectral structures, rather than the absolute photon number or readout contrast.

\end{document}